\documentclass[12pt,preprint]{aastex}
\usepackage{emulateapj5}
\usepackage{apjfonts}
\usepackage{psfig}
\shorttitle{Occultation and Microlensing}
\shortauthors{Agol}

\begin{document}


\title{Occultation and Microlensing}

\author{Eric Agol\altaffilmark{1,2}}
\altaffiltext{1}{Chandra Fellow}

\altaffiltext{2}{California Institute of Technology, Mail Code 130-33, 
Pasadena, CA 91125 USA; agol@tapir.caltech.edu}

\slugcomment{}


\begin{abstract}
Occultation and microlensing are different limits of the same phenomena
of one body passing in front of another body.  We derive a general
exact analytic expression which describes both microlensing and
occultation in the case of spherical bodies with a source of uniform 
brightness and a non-relativistic foreground body.  We also compute 
numerically the case of a source with quadratic limb-darkening.
In the limit that the gravitational deflection angle is comparable
to the angular size of the foreground body, both microlensing and
occultation occur as the objects align.  Such events may be used to constrain 
the size ratio of the lens and source stars, the limb-darkening 
coefficients of the source star, and the surface gravity of the lens 
star (if the lens and source distances are known).  Application of these 
results to microlensing during transits in binaries and giant-star 
microlensing are discussed.  These results unify the microlensing and 
occultation limits and should be useful for rapid model fitting of 
microlensing, eclipse, and ``microccultation'' events.
\end{abstract}



\section{Introduction}
When two stars (or other bodies) come into close alignment on the sky, the 
foreground star may either eclipse or microlens the background star.  As the 
stars align, if the angular size of the foreground star is much larger than 
its gravitational deflection angle, then the foreground star can eclipse; 
if the contrary is true then it can magnify.  More precisely, gravitational 
lensing by a point mass produces two images of a distant object, one interior
and one exterior to the Einstein radius in the lens plane,
$R_E=[4R_G D_L(D_S-D_L)/D_S]^{1/2}$ where $R_G=GM/c^2$ is the gravitational
radius for a lens of mass $M$, and $D_{L,S}$ are the distances to the 
lens or source.  Both images move toward the Einstein
radius as the lens and source approach, so the outer image will be 
occulted during the approach if the radius of the lens is larger 
than the Einstein radius.  The inner image, however, starts off 
near the origin and thus is occulted when the source is far from
the lens.  As the lens and source approach, the inner image can become 
unocculted if the lens is smaller than the Einstein radius (Figure 1).
Occultation is most important in microlensing if $R_E \sim R_L$,
where $R_L$ is the radius of the lens star (assumed to be spherical).
In Galactic microlensing, typically $R_L \ll R_E$, so the occultation
of the inner image occurs, but is usually rather faint.
In special circumstances, such as in eclipsing binaries containing 
compact objects \citep{mae73,mar01} 
or lensing by giant stars, $R_L \sim R_E$, 
so the effects of both microlensing and occultation must be included.
This ``microccultation'' can show more varied behavior than the
usual microlensing or occultation lightcurves and can be used
to constrain the surface gravity of the lens star \citep{bro96}.

\begin{figure*}
\centerline{\psfig{file=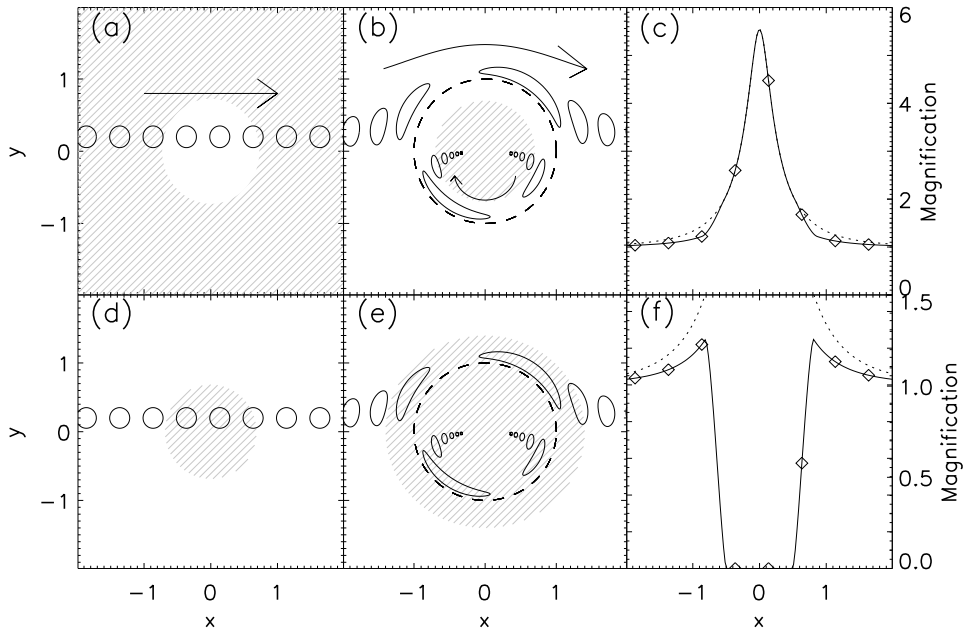,width=5in}} 
\figcaption[]{(a) Stars at various positions in source plane.  Shaded region
shows the area in which the inner image is occulted by the lens star for 
$r_L<1$, which is the region $\beta > \beta_L = 1/r_L -r_L$.  The axes have
units of $R_E D_S/D_L$.  (b) Images of the star in the image plane. 
The arrow is imaged as well for reference.  The dashed line is 
$R_E$.  (c) Magnification as a function of position - dotted line
is without occultation, while solid line includes occultation.  The symbols show
the magnification of the source at the positions depicted in (a).
(d)-(f) Same as (a)-(c), but for $r_L>1$.  In this case, the inner
image is fully occulted, and the shaded region in (d) shows where the
outer image is occulted by the lens star.}
\end{figure*}

\citet{mae73} and \citet{mar01} have carried out numerical computations of 
microccultation lightcurves.  Here we present an exact {\it analytic} solution for 
the lightcurve of a uniform source which agrees with their work, and we
present numerical calculations for limb-darkened sources. \citet{bro96} and
\citet{boz02} computed lightcurves for lensing events, treating the source as 
a point source, while the expressions presented here are valid for 
extended and limb-darkened sources as well.
In section 2 we discuss microlensing and occultation of a point source.
In section 3 we include the finite size of a uniform 
source.  In section 4 we include the effects of limb-darkening numerically
for microlensing and occultation.
In section 5 we apply the results to several astrophysical cases of 
possible interest, namely white dwarf-main sequence binaries, 
microlensing in globular clusters, and microlensing by supergiants.
In section 6 we summarize.

\section{Point source}

The lensing equation for a point source and point lens, neglecting diffraction
and strong relativistic effects, is \citep{sch92} 
\begin{equation}\label{lenseq}
\beta = \theta - {1 \over \theta},
\end{equation}
where $\theta$ and $\beta$ are the image and source position angles 
in units of $R_E/D_L$.
In the limit that the source is much smaller than the Einstein
radius and is not aligned with the lens, then the point-source
magnification is an adequate approximation \citep{pac86}. 
Solving this equation, the image positions are
\begin{equation}\label{pointlens}
\theta_\pm = {1 \over 2}\left[\beta \pm \sqrt{4 + \beta^2}\right],
\end{equation}
where $\theta_-$ is the image interior to the Einstein radius and 
$\theta_+$ is the image exterior to the Einstein radius 
($\left|\theta_-\right| < 1$ and $\theta_+ > 1$).  
The magnifications of the images are 
\begin{equation}
\mu^p_\pm = {1 \over 2} \left( {2 + \beta^2 \over \beta \sqrt{4 + \beta^2}}\pm 1 \right),
\end{equation}
or $\mu^p_+=(1-\theta_+^{-4})^{-1}$ and $\mu^p_-=(\theta_-^{-4}-1)^{-1}$.

If the size of the lens star is less than the size of the Einstein radius,
$r_L=R_L/R_E <1$, then the inner image will be occulted for
$\left|\theta_-\right| < r_L$.  This corresponds to $\beta > r_L^{-1} - r_L$, which 
means that the inner image is occulted when the source is distant
from the lens and unoccults when the source approaches the lens.
The change in magnitude at the point of the occultation of the 
inner image is
\begin{equation}
\Delta m = -2.5 \log{\left[1 + {r_L^4 \over 1+f(1-r_L^4)}\right]},
\end{equation}
where $f=F_L/F_S$ is the ratio of the flux from the lens to the
unlensed flux from the source. 
In typical Galactic microlensing
events, $r_L \ll 1$, so the inner image usually appears (unoccults)
when the source is distant from the lens, so the change in magnitude
is very small.  However, if the images can be resolved directly and
the demagnified inner image is brighter than the lens star, then
appearance of the inner image might be detectable.

If the size of the lens star is greater than the Einstein radius,
$r_L > 1$, then the inner image is always occulted and the outer 
image will be occulted for $\theta_+ < r_L$ which corresponds to
$\beta < r_L - r_L^{-1}$.  During occultation the source star
disappears so that one can only see the lens star.  Thus, the total
magnification for microlensing of a point source is
\begin{equation} \label{pointmag}
\mu^p = \mu^p_+\Theta\left(r_L\beta - r_L^2+1\right) + \mu^p_-\Theta\left(1-r_L^2-r_L\beta\right),
\end{equation}
where 
\begin{equation}
\Theta(x)=\left\{\begin{array}{ll}
1 & x > 0\\
0 & x \le 0\\
\end{array}\right.
\end{equation} 
is the step-function.
In the limit $r_L=0$, this reduces to the usual microlensing magnification
\citep{pac86}, 
while in the limit $r_L\gg 1$, $\mu^p_-$ is negligible, $\mu^p_+\sim 1$, and
occultation simply occurs when $\beta < r_L$.

At the point of occultation, $\beta=\pm\left(r_L^{-1}-r_L\right)$,
and since $\beta$ can be measured from a fit to the microlensing lightcurve, 
one can measure $r_L$ \citep{bro96}.  The sign is determined by
whether the image appears ($+$) or disappears ($-$) at the center of the
event.

The average astrometric position of the images during the microlensing 
event is \citep{wal95} 
\begin{equation}
\Delta \theta = {\mu^p_+ \theta_+ + \mu^p_-  \theta_- \over  \mu^p + f},
\end{equation}
where the difference in image position is measured with respect to the position of
the lens on the sky.  When $r_L \ll 1$, then the change in position during
occultation of the inner image is 
\begin{equation}
\Delta \theta (r_L) = -{r_L^3 \over (1 + f)^2}.
\end{equation}
This has a weaker dependence on $r_L$ than the magnitude change, but will
still be quite small unless $r_L \sim 1$.
In the case where $r_L \gg 1$, then the source will be completely occulted, so
the centroid change is simply
\begin{equation}
\Delta \theta (r_L) = {r_L \over f+1}.
\end{equation}

The point source approximation has two limitations:  during eclipse
ingress or egress, the finite size of the source causes a smooth
transition, and if the surface brightness of the source and lens
are similar and $r_L\sim 1$, then the source must also have have
a size similar to the Einstein radius to contribute a significant fraction
of the flux.  Thus, in the next section we derive a more general
formula including the finite extent of the source.


\section{Extended uniform source}

In the case of a circular source with uniform surface brightness, the
images may be partly or fully eclipsed by the lens star.  The 
magnification (or dimming) is equal to the ratio of the unocculted area of the 
images to the area of the unlensed source since surface brightness
is conserved during lensing.  For a uniform
source the area can be computed by integrating over the image
boundaries using Stokes' theorem \citep{gou97,dom98}. 
In the case of a point-mass lens, this integral can be solved
analytically as first shown by \citet{wit94} 
for $r_L=0$.  For an extended source with normalized radius 
\begin{equation}
r_S={R_SD_L\over R_E D_S},
\end{equation}
it is more useful to define a two-dimensional lensing equation to
integrate over the source.  \citet{wit94} 
define a complex lensing equation 
\begin{equation}\label{complex}
\zeta=z-{1 \over \bar z},
\end{equation}
where $\zeta$ is the complex coordinate for the source plane in units of $R_E D_S/D_L$ and
$z=x+iy$ is the complex coordinate for the lens plane in units of $R_E$.   Comparing
to equation \ref{lenseq}, $|\beta|=|\zeta|$ and $|\theta|= |z|$.

We assume that the source has a uniform surface brightness in
the region $\zeta_0+re^{i\phi}$ ($\zeta_0$ is real and positive) where 
$0\le \phi < 2\pi$ and
$0\le r \le r_S$, while we assume that the lens is opaque in
the region $0\le |z| \le r_L$.  The solution for the image
positions is
\begin{equation}
z_\pm = {\zeta \over 2}\left[1 \pm \sqrt{1+{4 \over \zeta \bar \zeta}}\right],
\end{equation}
which is the complex version of equation \ref{pointlens}.
The lensing magnification for a uniform source is simply the ratio of 
the area of the lensed images to the area of the source.  The
integral over area can be converted to an integral over the source
boundary using Stokes' theorem \citep{gou97}, 
giving
\begin{equation}\label{stoke}
\mu_\pm=\pm{1 \over \pi r_S^2}\int d\phi |z_\pm|^2{\partial \Psi \over \partial \phi},
\end{equation}
where $\Psi=\cot^{-1}\left({\zeta_0r_S^{-1}\csc{\phi} + \cot{\phi}}\right)$ is the
position angle of the image.  The total magnification is
\begin{equation} \label{totmag}
\mu=\mu_+ +\mu_-.
\end{equation}
In case of a finite-sized lens, 
$z_\pm$ should be replaced in the integrand with $r_Le^{i\Psi}$ whenever 
$|z_\pm| \le r_L$.  In other words, when an image is partially occulted,
then the inner boundary is given by the edge of the lens, while the outer
boundary is given by the edge of the outer image.

There are several different cases to consider:
\begin{enumerate} 
\item$r_S=0$: Point source (equation \ref{pointmag}); 
\item$r_S>0$, $r_L=0$:    Extended source, unocculted \citep{wit94}; 
\item$r_S>0$, $1>r_L>0$:   Extended source, inner image may be partly occulted,
outer image unocculted;
\item$r_S>0$, $r_L>1$:  Extended source, inner image fully occulted,
outer image may be partly occulted.
\end{enumerate} 
The expressions for $\mu_\pm$  in each of these cases are summarized in 
Tables 1 and 2, where $p, e,$ and $o$ superscripts refer to the
point-source magnification \citep{pac86}, 
extended-source magnification \citep{wit94}, 
or occulted-extended source magnification (below),
respectively and $\beta_L=|r_L^{-1}-r_L|$.  Each of the magnification expressions 
in Tables 1 and 2 are given in equations \ref{pointmag} and 
\ref{firsteq}-\ref{lasteq},  with the the range of the variables
for which the functions apply given in the columns.

As an example of how the computation proceeds, we consider the case in 
which $r_L>1$ and the source overlaps
the shadow of the lens (for example, the source in Figure 1(d) at $x=0.6$).
In this case the inner image is completely occulted while the outer image 
is partially occulted, so we must integrate equation \ref{stoke} for
$\phi$ between $\pm\chi=\pm\cos^{-1}\left[\left(\beta_L^2-r_S^2-\zeta_0^2\right)
/\left(2r_S\zeta_0\right)\right]$.
This gives an area which is between the outer edge of the source image and 
origin, so we need to subtract off the area within the shadow which is
$r_L^2\phi_2=r_L^2\cos^{-1}\left[\left(r_S^2-\beta_L^2-\zeta_0^2\right)/\left( 
2 \beta_L \zeta_0\right)\right]$.  Thus,
\begin{equation}
\mu^o_+={1 \over \pi r_S^2}\int_{-\chi}^{\chi}
d\phi |z_\pm|^2{\partial \Psi \over \partial \phi} - {\phi_2 r_L^2\over
\pi r_S^2}.
\end{equation}
Making the substitution $u=\beta^2=\zeta_0^2+r_S^2+2\zeta_0 r_S \cos{\phi}$,
this equation is
\begin{equation}
\mu^o_+ = {1 \over 4\pi r_S^2}\int_{\beta_L^2}^{u_2}du
{(u-u_3)\left(1+2u^{-1} +\sqrt{4+u\over u}\right)
\over \left[(u_2-u)(u-u_1)\right]^{1/2}} - {\phi_2 r_L^2
\over \pi r_S^2},
\end{equation}
where $u_1$, $u_2$, and $u_3$ are defined below.  The integral
can be reduced to elliptic integrals as given below.

For an extended source ($r_S > 0$), the magnification of the inner image
when $1>r_L>0$ is given by:
\begin{eqnarray}\label{firsteq} 
\mu^o_- &=& {\Theta(r_S-\zeta_0)\over r_S^2} -{1 \over 4 \pi r_S^2} 
\left[\vphantom{\left(\sqrt{1+4/u_0}-1\right)}
2(1+r_S^2) \phi_1 -4 sgn(u_3) \phi_0 + 4 r_L^2 \phi_2   \right. \cr
&+&\left. \sqrt{(u_2-u_0)(u_0-u_1)} \left(\sqrt{1+4/u_0}-1\right)
- G(\phi_0)   \right],
\end{eqnarray}
where
\begin{eqnarray}
G(\phi) &=& {1\over \sqrt{u_2(4+u_1)}}\left[
\vphantom{r_S^2}
u_2(4+u_1)E(\phi,k_1) -(u_1u_2+8u_3)F(\phi,k_1)\right.\cr
&+&\left.4u_1(1+r_S^2)\Pi(\phi,n,k_1)\right],
\end{eqnarray}
$F, E,$ and $\Pi$ are elliptic integrals of the first, second and third
kinds \citep{gra94}, 
 $sgn(x)$ chooses the
sign of $x$, and the other variables are
\begin{eqnarray}\label{othervar}
u_0&=&\beta_L^2 \cr
u_1&=&(\zeta_0-r_S)^2 \cr
u_2&=&(\zeta_0+r_S)^2 \cr
u_3&=&\zeta_0^2-r_S^2 \cr
\phi_0 &=& \cos^{-1}{\sqrt{u_1(u_2-u_0)\over u_0(u_2-u_1)}}\cr
\phi_1 &=& \cos^{-1}{\left[u_1+u_2-2u_0 \over u_2-u_1\right]}\cr
\phi_2 &=& \cos^{-1}{\left[u_3+u_0 \over 2\zeta_0 u_0^{1/2}\right]}\cr
n      &=& 1-{u_1\over u_2} \cr
k_1^2    &=& {4(u_2-u_1)\over u_2(4+u_1)}.
\end{eqnarray}

In the special case that $\zeta_0=r_S$, the magnification of the
inner image becomes
\begin{eqnarray} 
\mu_-^{o,*} &=& \mu_-^{e,*} +{1\over \pi r_S^2} \left[\vphantom{\beta_L\over 2r_S}
\left(1+r_S^2-r_L^2\right) \cos^{-1}{\beta_L \over 2r_S}+{v_2 \over 4}(\beta_L-v_1) \right. \cr
&-& \left.
(1+r_S^2)\tan^{-1}{v_2 \over v_1}\right]
\end{eqnarray}
where
\begin{eqnarray}
v_1&=&\sqrt{4+\beta_L^2} \cr
v_2&=&\sqrt{4r_S^2-\beta_L^2},
\end{eqnarray}
and
\begin{equation}
\mu_\pm^{e,*} = {1 \over \pi r_S^2}\left[ r_S +(1+r_S)^2 \tan^{-1}{r_S}\right] \pm {1\over 2}.
\end{equation}
which is the magnification of the unocculted images when $\zeta_0=r_S$.

\centerline{\psfig{file=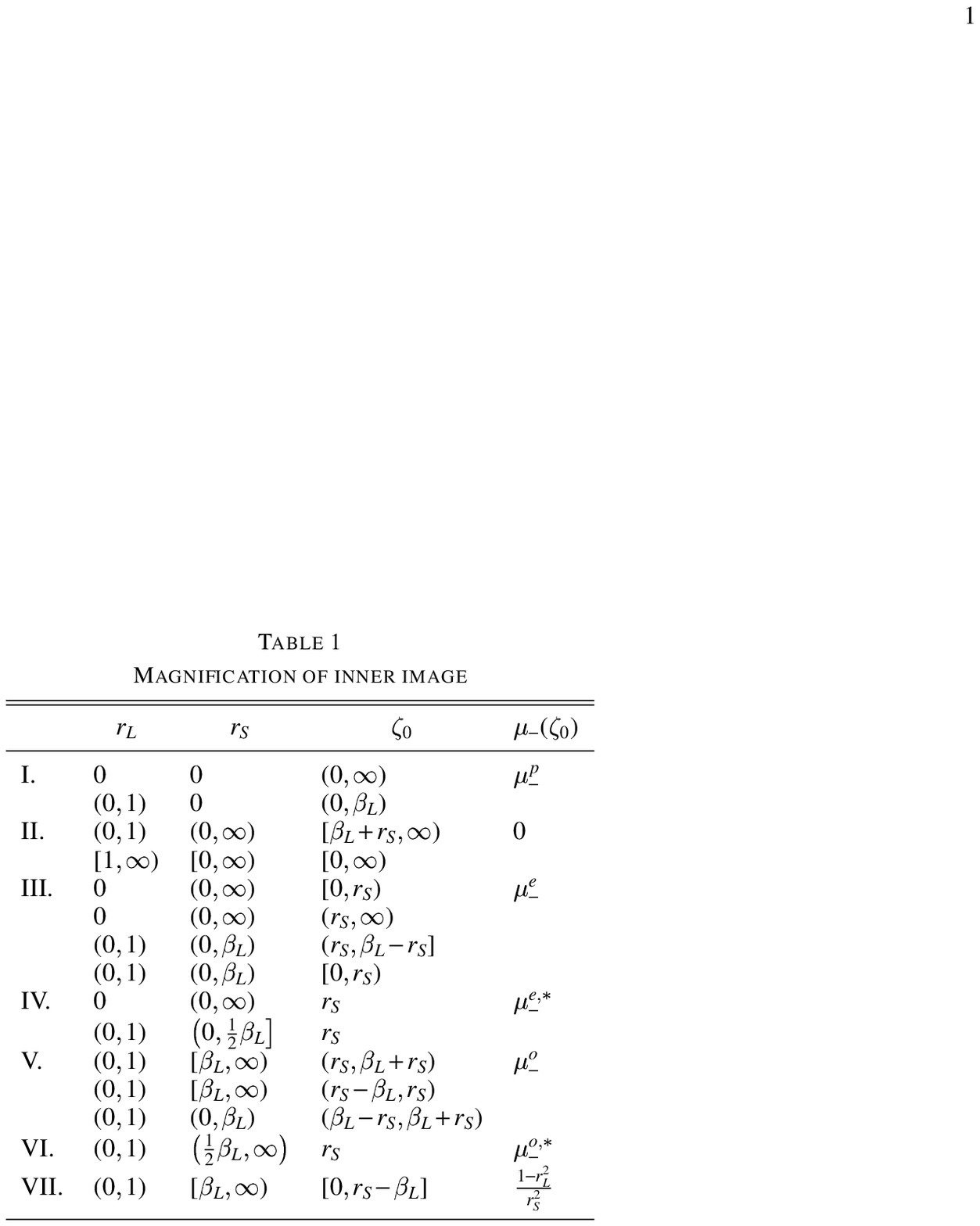,width=\hsize}} 

When $r_L > 1$, then the outer image can be occulted.  The magnification
in this case is given by:
\begin{eqnarray}
\mu_+^o&=&{1\over 4\pi r_S^2}\left[2(1+r_S^2)\psi_2-4sgn(u_3)\psi_1-4r_L^2\phi_2  \right.\cr
&+&\left.\sqrt{(u_2-u_0)(u_0-u_1)}\left(\sqrt{u_0/(4+u_0)}+1\right)+G(\psi_0)\right],
\end{eqnarray}
where
\begin{eqnarray}
\psi_0&=&\cos^{-1}{\sqrt{(u_0-u_1)(4+u_2)\over (4+u_0)(u_2-u_1)}},\cr
\psi_1&=&{\pi \over 2}-\phi_0, \cr
\psi_2&=&\pi-\phi_1+2
\cos^{-1}\sqrt{u_0(4+u_0)\over u_0(4+u_1+u_2)-u_1u_2},
\end{eqnarray}
and the other variables are as in equation \ref{othervar}.
\centerline{\psfig{file=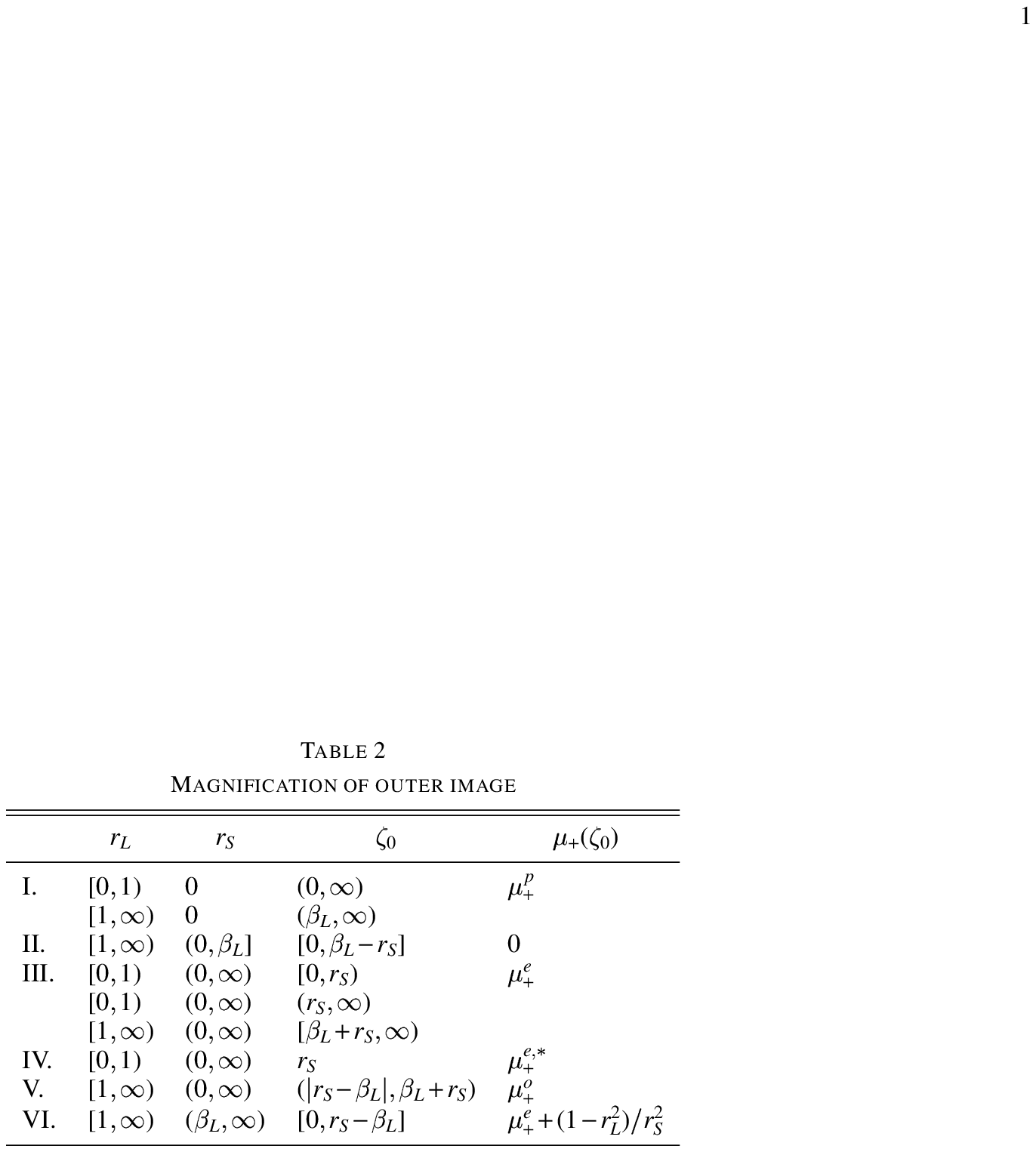,width=\hsize}} 

When the inner or outer images are unocculted, then the magnification is 
\begin{eqnarray} \label{lasteq}
\mu^e_-&=&{G(\pi/2)\over 4\pi r_S^2}- {1\over 2},\cr
\mu^e_+&=&{G(\pi/2)\over 4\pi r_S^2}+ {1\over 2},
\end{eqnarray} which 
agrees with the expression of \citep{wit94}. 
In principle, one could also compute the image centroid for the
source including occultation \citep[as done by][]{wit95}.

We now provide several graphical examples of these equations.
Figure 2 shows the magnification for a source with $r_L=0.9$ and
$r_S=0.25$, compared with cases in which either $r_L=0$ or $r_S=0$.  
In all three cases the outer image is unobscured, but
the inner image appears when the source approaches the lens.  In the
point source case, the appearance is abrupt and creates a strong
brightening, while for the extended source the appearance is more gradual.

\centerline{\psfig{file=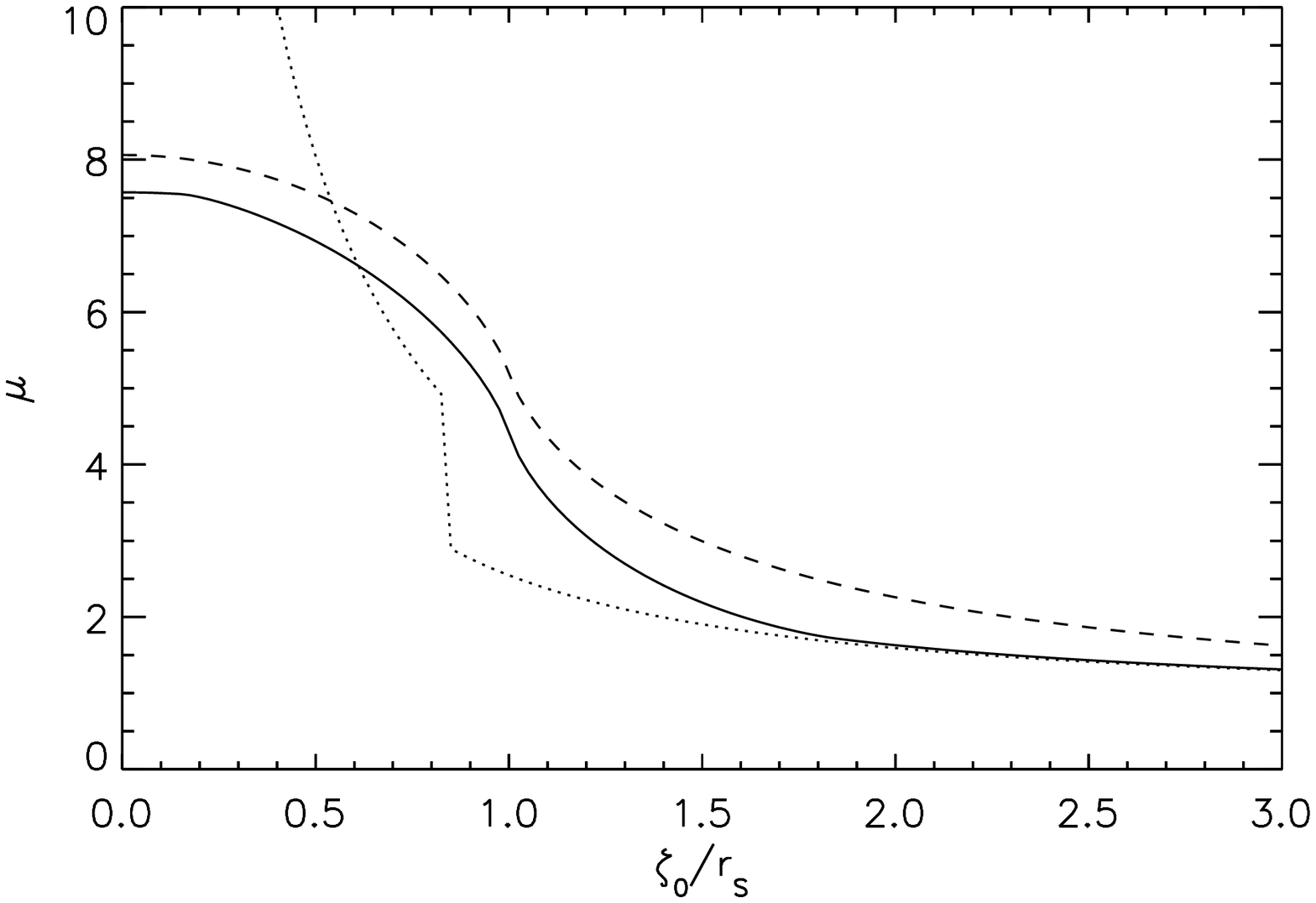,width=\hsize}} 
\begin{normalsize}
{\textsc{Fig}.~2. --  Magnification for $r_S=0.25$, $r_L=0.9$ (solid),
$r_S=0$, $r_L=0.9$ (dotted - in this case the horizontal axis is
scaled by $r_S=0.25$ for comparison), $r_S=0.25$, $r_L=0$ (dashed).}
\end{normalsize}

Figure 3 shows the magnification for a source with $r_L=1.1$ and
$r_S=0.25$, compared with cases in which either $r_L=0$ or $r_S=0$.  
The finite size of the lens and the source creates both magnification
and occultation, but the magnification wins out ($\mu>1$) in this
case. In all three cases the inner image is obscured, while the outer
image can be obscured when the source approaches the lens.  In the
point source case the occultation is abrupt and results in disappearance
of the background source. 

\centerline{\psfig{file=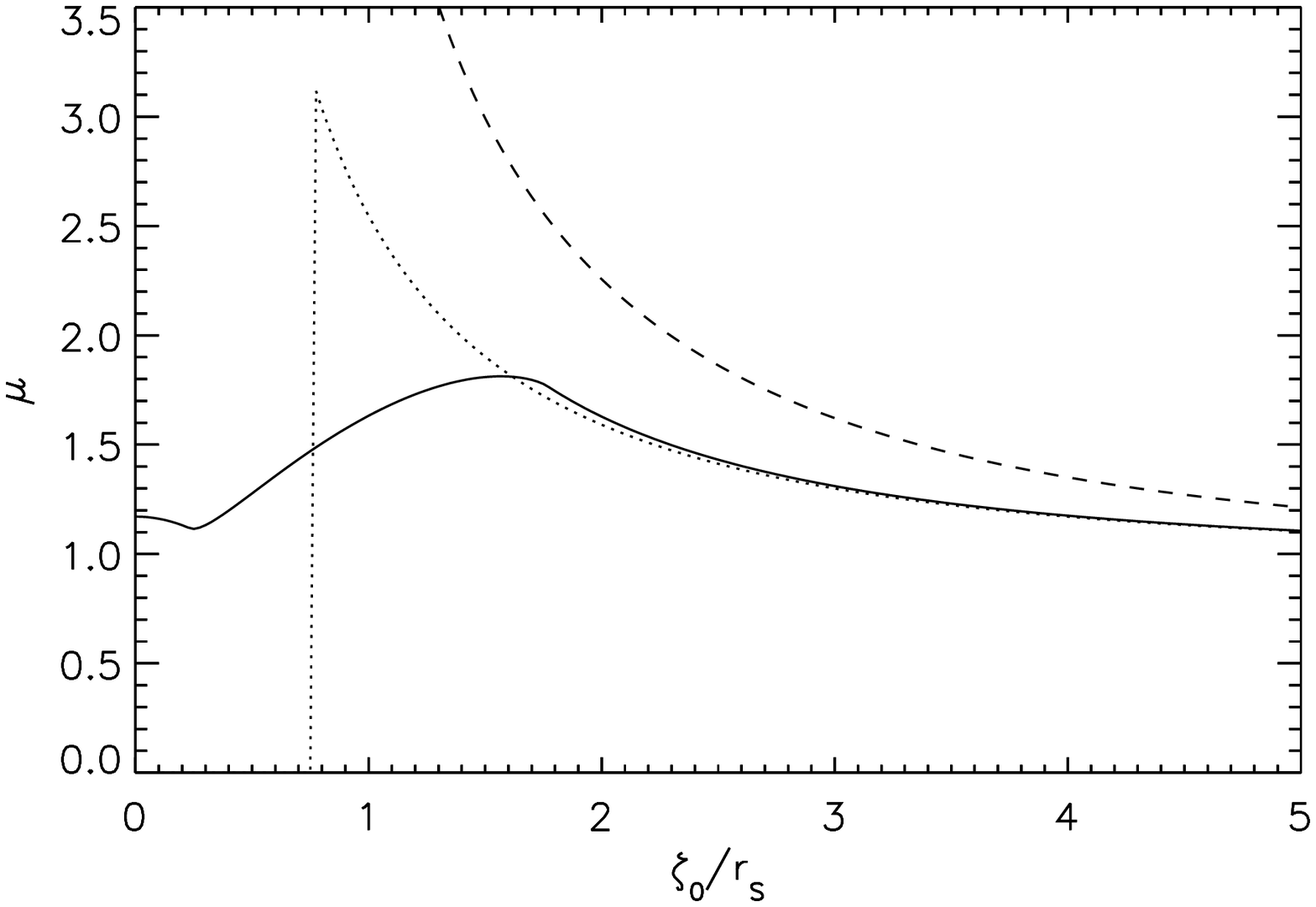,width=\hsize}} 
\begin{normalsize}
{\textsc{Fig}.~3. -- Magnification for $r_S=0.25$, $r_L=0.9$ (solid),
$r_S=0$, $r_L=0.9$ (dotted - in this case the horizontal axis is
scaled by $r_S=0.25$ for comparison), $r_S=0.25$, $r_L=0$ (dashed).}
\end{normalsize}

For the case of $r_S=1$, we show several different values of $r_L$
in Figure 4.   In all cases with finite $r_L$ the magnification shows 
a much flatter profile near the origin than for the $r_L=0$ case.
In the limit $r_L \ll 1$ the lightcurve approaches that of
extended-source microlensing, while for $r_L \gg 1$ approaches
the limit of occultation.  

\centerline{\psfig{file=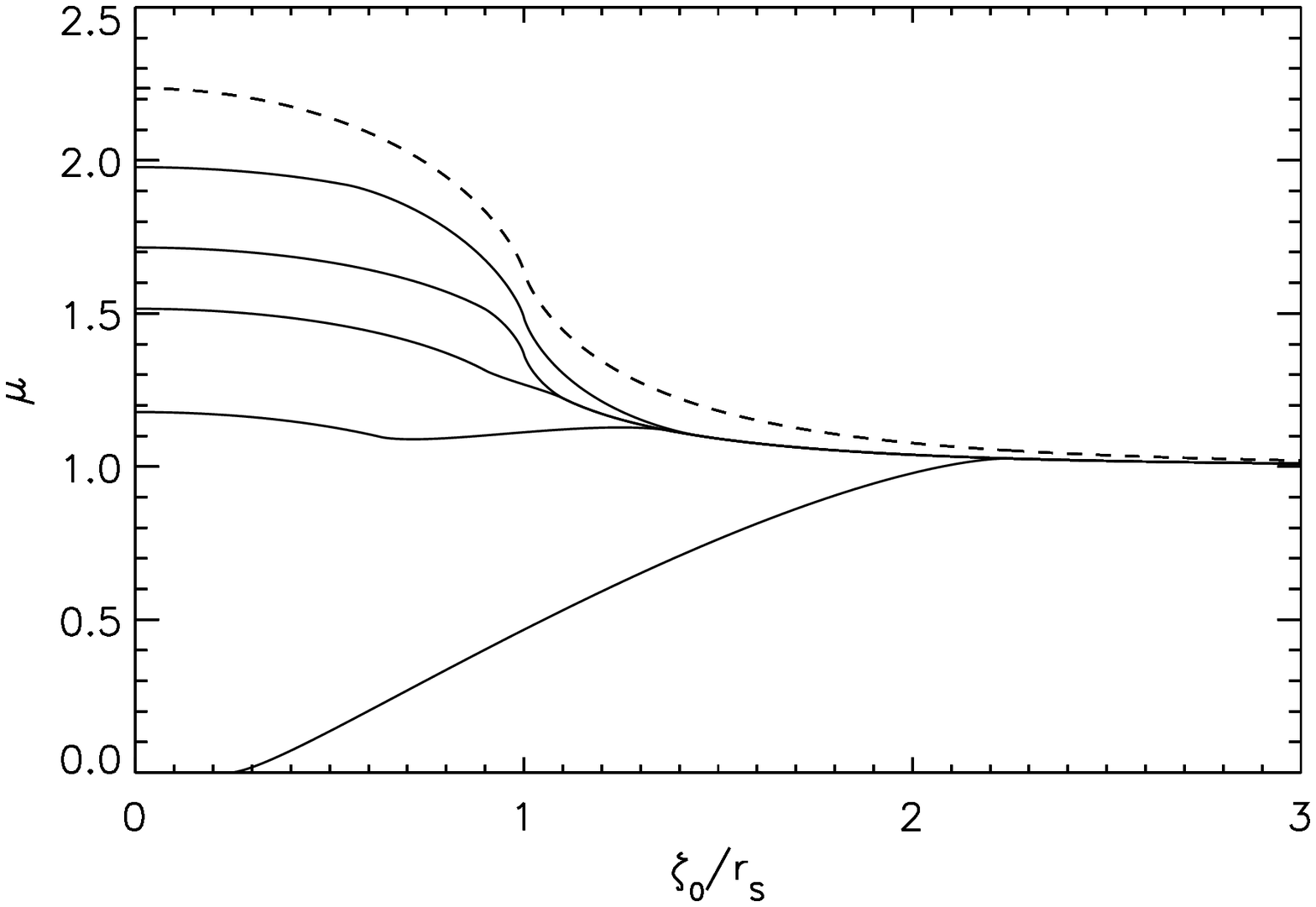,width=\hsize}} 
\begin{normalsize}
{\textsc{Fig}.~4. --  Magnification for $r_S=1$ and $r_L=0.8,0.95,1.05,1.2,1.8$
(solid lines from top to bottom); $r_L=0$ is dashed line.} 
\end{normalsize}

Figure 5 shows magnification for different source sizes, but
fixed lens size $r_L=0.95$.  The smallest sources show broad sloping
wings indicative of the appearance of the second image; in the smallest
case the source becomes completely revealed near the origin.  For
the largest cases the magnification shows a sharper slope near
$\zeta_0=r_S$ than in the $r_L=0$ case.

\centerline{\psfig{file=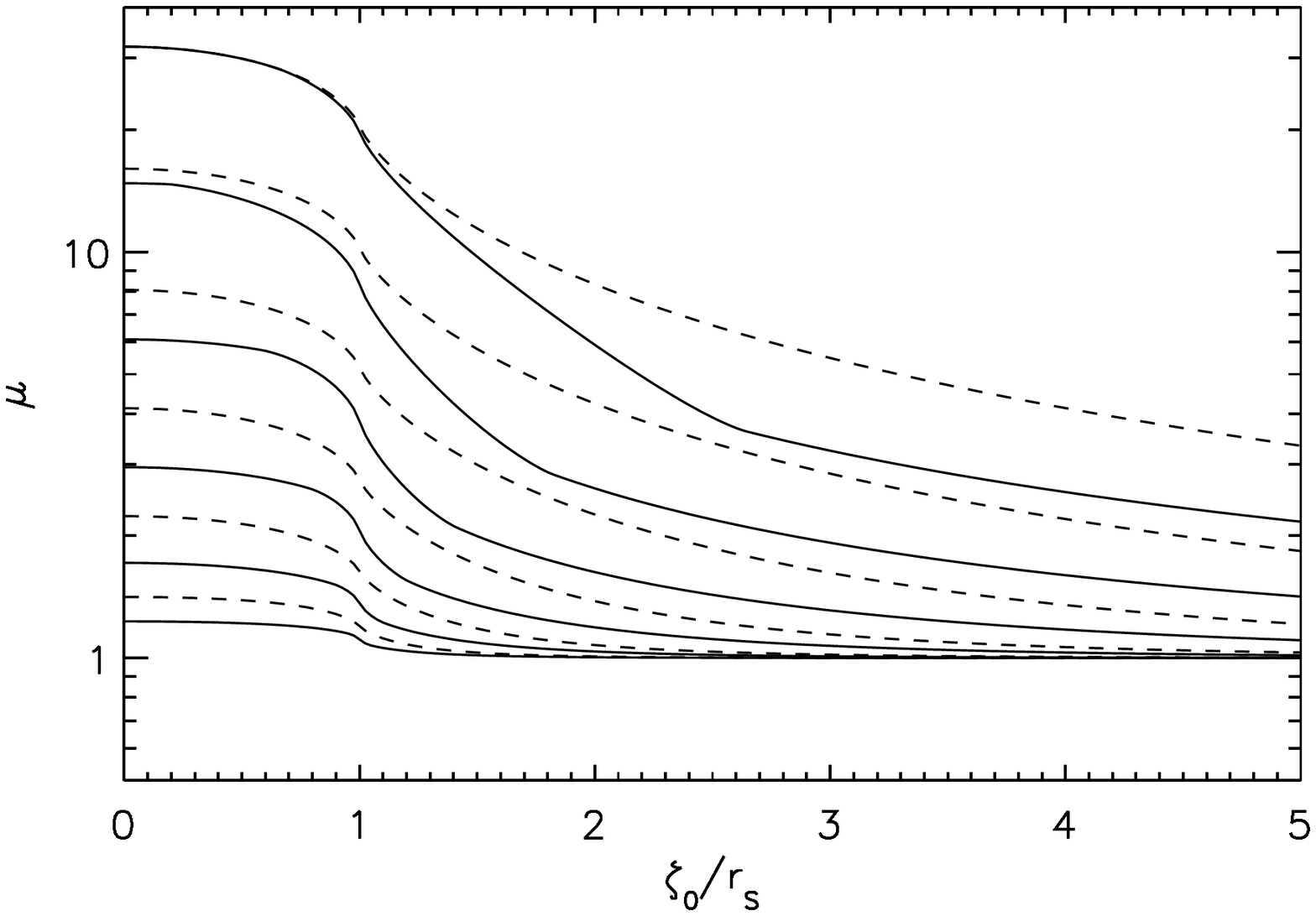,width=\hsize}} 
\begin{normalsize}
{\textsc{Fig}.~5. --  Magnification for $r_S={1\over 16},{1\over 8},{1\over 4},{1\over 2},1,2$
from top to bottom. Solid lines show $r_L=0.95$, while dashed lines
show $r_L=0$.} 
\end{normalsize}

A uniform source causes rather sharp features in the lightcurve during
the ingress and egress of the occultation, and leads to flatter
lightcurves during transit.  However, a limb-darkened source has
a smoother ingress/egress and has curvature during a transit.  Thus, in
the next section we consider microlensing and occultation of a limb-darkened
source.

\section{Limb Darkening}

Limb darkening causes a star to be more centrally peaked in brightness compared
to a uniform source.  This leads to larger magnification during microlensing or
larger dimming during transit/occultation.  Thus, including limb-darkening
is important for computing accurate microlensing/occultation lightcurves.
Describing limb-darkening with a quadratic law
\begin{eqnarray}
{I(r)\over I(0)} &=& 1-\gamma_1(1-\sigma)-\gamma_2(1-\sigma)^2,\cr
\sigma&=&\sqrt{1-\left({r\over r_S}\right)^2},
\end{eqnarray}
where $\gamma_1+\gamma_2 < 1$,
leads to a magnification  of
\begin{equation} \label{mulimb}
\mu(r_L,r_S,\zeta_0,\gamma_1,\gamma_2) = \left[\int_0^{r_S} dr I(r)\right]^{-1} \int_0^{r_S} dr I(r) {d \mu r^2 \over dr},
\end{equation}
where $\mu(r)$ can be computed from the expressions in the previous section
(replacing $r_S$ with $r$).
We could use a more accurate limb-darkening formula, but rely on
a quadratic law for simplicity.
Given the complicated dependence of the magnification on the radius, this 
integral is best done numerically using a finite-difference approximation for 
the derivative of the uniform
magnification.  An example is shown in Figure 6 - in this case the dip during
occultation is deeper due to limb-darkening since the source is brighter at
the center, and thus more flux is lost, and the magnification
decreases toward the origin rather than increasing as in the uniform source
case.  Both the uniform and the limb-darkened cases are shallower
when compared to the pure occultation case due to magnification of the background
source.  A second example is shown in Figure 7. In this case the limb-darkening
causes a weaker magnification as the outer limb is magnified, while the peak 
is increased due to the more concentrated brightness.  In the special case that 
$\zeta_0=0$, the integral is tractable analytically as follows, and is shown as 
the solid dots in Figures 6 and 7.

\centerline{\psfig{file=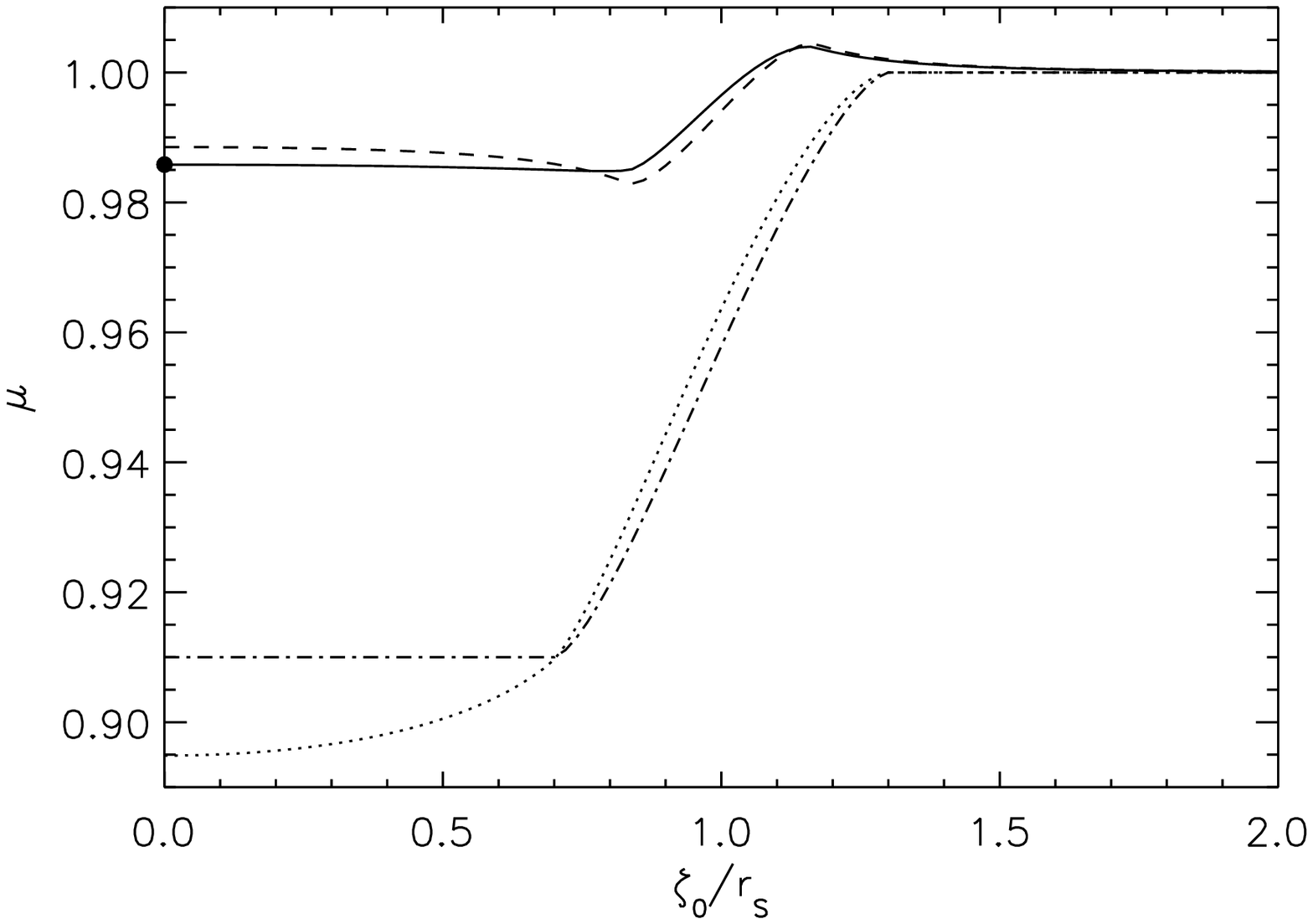,width=\hsize}} 
\begin{normalsize}
{\textsc{Fig}.~6. -- Magnification for $r_S=5$ and $r_L=1.5$.  The solid
line is limb-darkened with $\gamma_1=\gamma_2=0.3$ (equation \ref{mulimb}), 
while the dashed line is a uniform source (equation \ref{totmag}, Tables
1 and 2).  The dash-dot line shows a uniform source and lens of the same size ratio but
neglecting lensing, while the dotted line is for a limb-darkened source neglecting lensing.}
\end{normalsize}

For $r_L>0$ and $r_S > \beta_L$, the magnification for a limb-darkened 
source at $\zeta_0=0$ becomes
\begin{eqnarray} \label{limbmag1}
\mu &=& {\sqrt{4+r_S^2}\over 6\Omega r_S^3}\left[\alpha_1\left((2+r_S^2)E(\phi_3,k_2)
-2F(\phi_3,k_2)\right) +\alpha_2 r_S\right] \cr
&+&{1\over 6\Omega r_S^3}\left(s \beta_L(4+\beta_L^2)^{1\over 2}+\alpha_3\right)
\left(\alpha_1\sqrt{\alpha_3}+\alpha_2-{3\gamma_2\over 2r_S}\alpha_3\right)\cr
&+&{2\gamma_2\over r_S^4\Omega}\left[{\alpha_3\over 8}(2+r_S^2) 
+\sinh^{-1}\left({r_S\over 2}\right)+\sinh^{-1}\left({s\beta_L\over 2}\right)\right],
\end{eqnarray}
where
\begin{eqnarray} \label{limbmag2}
\alpha_1&=&2(\gamma_1+2\gamma_2),\cr
\alpha_2&=&3(1-\gamma_1-\gamma_2)r_S-{3\gamma_2\over 2r_S}(2+r_S^2),\cr
\alpha_3&=&r_S^2-\beta_L^2,\cr
\Omega&=&1-\gamma_1/3-\gamma_2/6,\cr
\phi_3&=&\cos^{-1}\left(-s{\beta_L \over r_S}\right), \cr
s&=&sgn(1-r_L),\cr
k_2^2&=&{r_S^2 \over 4+r_S^2}.
\end{eqnarray}
When $\phi_3 > \pi/2$, then $E(\phi_3,k_2)=2E(k_2)-E(\pi-\phi_3,k_2)$ and
$F(\phi_3,k_2)=2K(k_2)-F(\pi-\phi_3,k_2)$.
For $r_L<1$ and $r_S<\beta_L$, the inner image is unocculted (the latter is always true for $r_L=0$),
and $\beta_L$ should be replaced by $r_S$ in equations \ref{limbmag1} and \ref{limbmag2} to give
\begin{equation}\label{limbmag3}
\mu={1\over \Omega r_S^2}\left[{\alpha_1\over 3k_2}\left((2+r_S^2)E(k_2)-2K(k_2)\right)+
{\alpha_2r_S\over 3k_2}
+{4\gamma_2\over r_S^2}\sinh^{-1}\left({r_S\over 2}\right)\right].
\end{equation}
For $\gamma_2=0$, this expression agrees with equation (A6) in \citet{wit95}. 

\centerline{\psfig{file=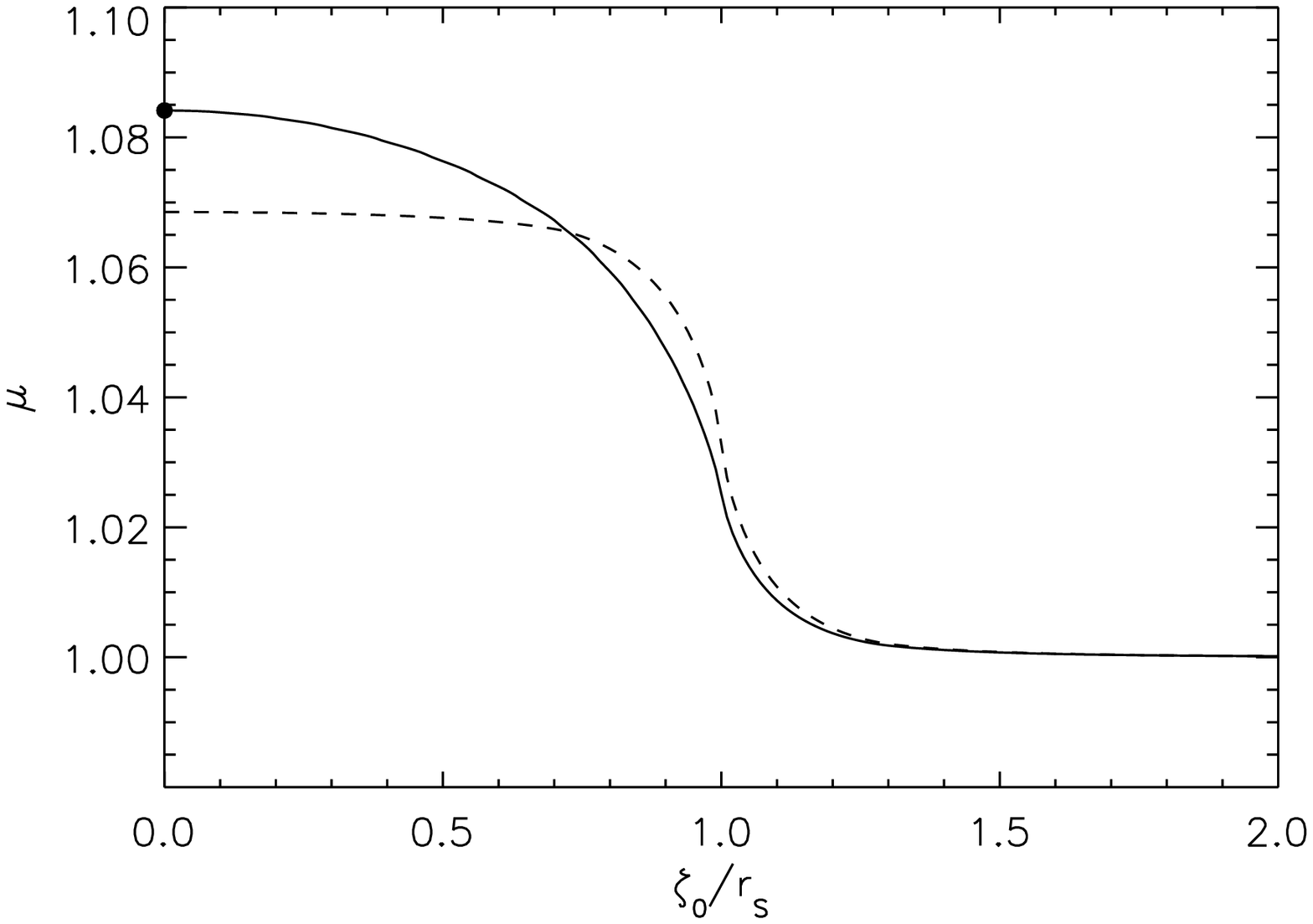,width=\hsize}} 
\begin{normalsize}
{\textsc{Fig}.~7. --  Magnification for $r_S=5$ and $r_L=0.5$.  The solid
line is limb-darkened with $\gamma_1=\gamma_2=0.3$, while the dashed line is a uniform
source.}
\end{normalsize}

In the occultation limit when the Einstein radius is small, $R_E \ll R_L, R_S$, then the 
lightcurve can be described by occultation only.  We can include
limb-darkening exactly in this case \citep{man02}.  An example is
shown in Figure 6 of the difference between occulting and microcculting lightcurves.

\section{Discussion}

The equations for microccultation are most relevant for
equality of the Einstein radius and lens radius which occurs if 
$D \equiv D_L(D_S-D_L)/D_S = R_L^2/(4 R_G)$.  We compute $D$
for several interesting objects in Table 3 (not to be confused with
their actual distances), including a white dwarf (Sirius B),
brown dwarf (Gliese 229b), red giant (Capella), blue supergiant (Rigel), 
yellow supergiant (Deneb), and red supergiant (Betelgeuse).
Application of the microccultation equations to white dwarfs,
brown-dwarfs, nearby supergiants, and giants in globular clusters
are discussed next.

White dwarfs in eclipsing binaries are the most likely location
to see microccultation \citep{mae73,mar01}.
Known white-dwarf binaries which transit
their companions have small semi-major axes (likely a selection effect), 
and thus small Einstein radii, changing the depth of transit by
only a few percent \citep{mar01}.  Due to common-envelope evolution,
white-dwarf binaries have typical semi-major axes $a\sim 0.1$ AU and masses
of $0.5 M_\odot$, so $R_E \sim 7\times 10^8 {\rm cm}(M/0.5 M_\odot)^{1/2}
(a/0.1 {\rm AU})^{1/2}$.  This is comparable to the size of white
dwarfs, $\sim 10^9$ cm, so both occultation and microlensing will
be important.  \citet{far02} apply the formulae
derived here to estimate how many white dwarfs may be found in
transit searches for extrasolar planets.  For example, the Kepler 
survey \citep{koc98} may find $\sim 10-100$ white dwarfs, comparable
to the expected number of terrestrial planets.  White dwarf transit
events will require including both lensing and occultation in modeling 
the lightcurves.

\centerline{\psfig{file=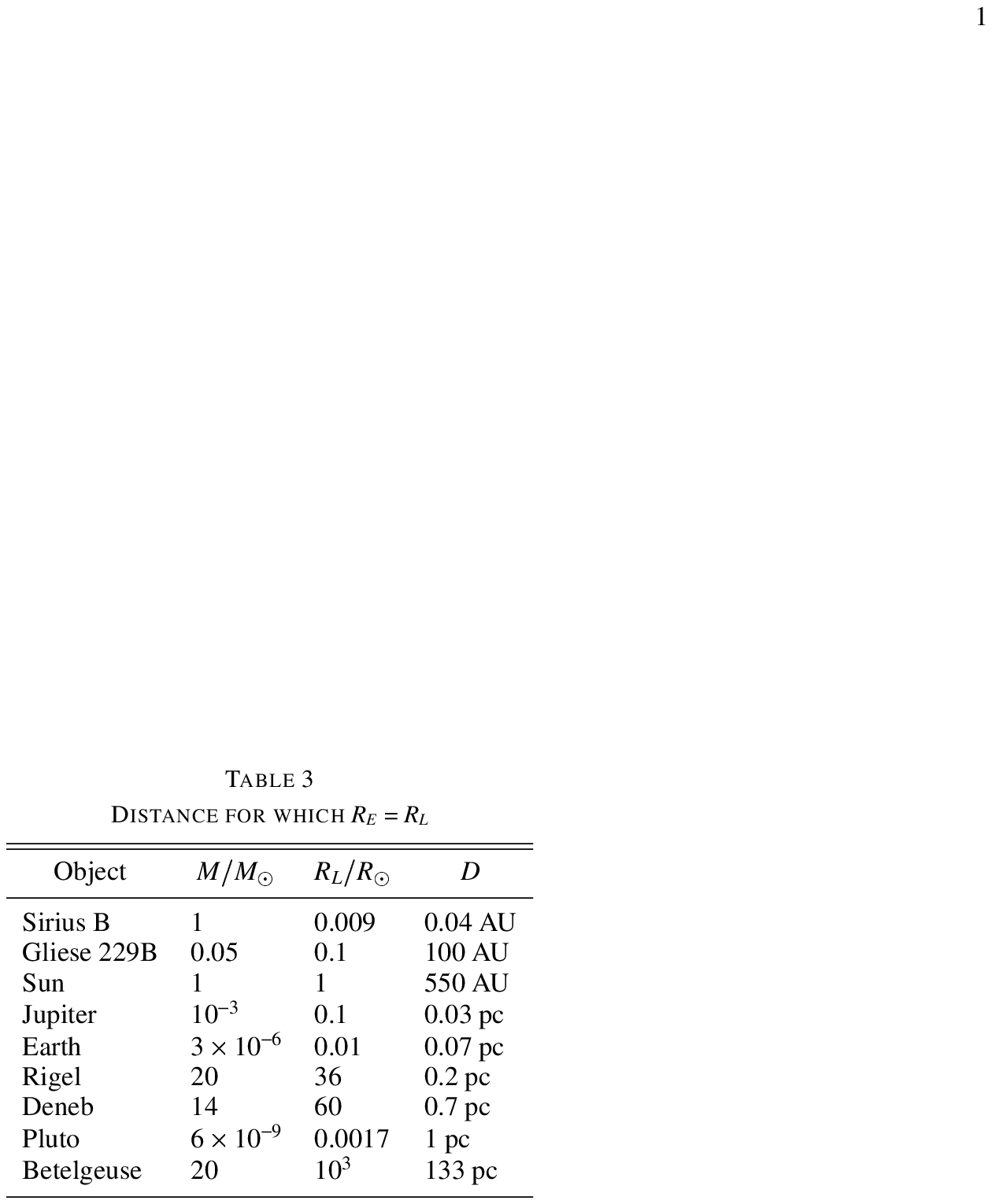,width=2.8in}} 

Though smaller in mass, brown dwarfs in eclipsing binaries
may have some observable microlensing effects during eclipse.
As gas planets and brown dwarfs have very similar sizes, their
transits of companion stars may look quite similar.  However, 
in the limit of a large source the depth of the
transit scales as $1+2(R_E/R_S)^{2}-(R_L/R_S)^2$ (for a uniform
source), so brown dwarfs will have a transit of smaller depth 
since $R_E$ is larger.  This will affect the measurement of 
limb-darkening which also changes the depth of the transit
(equation \ref{limbmag1}).  In the case of a 0.05 $M_\odot$
brown dwarf with radius $0.1 R_\odot$ in orbit at 1 AU about a G-type 
star, $r_L=10$ and $r_S=10^2$.  Thus, the transit depth differs 
by $2\times 10^{-4}$ from a $10^{-3} M_\odot$ planet of the same 
radius.  Such photometric precision can be obtained with
HST \citep{bro01} and other planned satellites, and will be indicated 
by a slight brightening outside of the transit (Figure 6).  The 
difference can be much larger, $\sim 1$\%, if the primary is also 
a brown dwarf.  

Lensing events caused by nearby stars may also show the signs
of both microlensing and occultation.  The most extreme case
is the star Betelgeuse which has a distance of $\sim 125$ pc and a mass
of $\sim 20 M_\odot$, giving an Einstein radius of $7 \times 10^{13}$
cm for sources at much larger distance.  This is only slightly
larger than the size of Betelgeuse, $\sim 4 \times 10^{13}$ cm,
so that distant stars passing behind Betelgeuse would create two
visible images as they approach;  a full occultation would never
occur (unless the mass of Betelgeuse were much smaller).  Distant
galaxies would create an Einstein ring surrounding the star with
the center occulted.  The challenge involved in carrying out
such an observation is resolving a faint background source from
the bright foreground star and waiting long enough for Betelgeuse
to pass in front of a star or galaxy.  In every microlensing
event an occultation should occur when $\beta=\beta_L$.  Since
this usually leads to a demagnified image, extremely accurate
photometry is necessary to see an occultation event.  

A final application of the microccultation equations is to giant 
stars acting as lenses in globular clusters.  There are about 
$3\times 10^5$ evolved giant stars in Milky Way globular clusters.
For a giant star at the clump with a mass of $1 M_\odot$, the Einstein
radius for a separation of 1 pc is $\sim 20 R_\odot$, comparable
to the size of the star, $\sim 20 R_\odot$.  If the relative velocity
is $\sim 10$ km/s then red giants in globular clusters cover
about $3\times 10^{31}$ cm$^2$ yr$^{-1}$ which is about $10^{-8}$
of the total area in globular clusters.  Thus, about $10^8$ stars in
globular clusters must be monitored to find a single red giant
transit event per year and a typical event will last about one month.  
A lightcurve of a lensing event by a red giant in a globular cluster
would allow one to measure $r_L$ and $r_S$, as well as $\gamma_1$ and
$\gamma_2$ for the source star.  Since the Einstein radius 
in this case is $R_E=\left[4 GM c^{-2} D\right]^{1/2}$ where $D$ is the
separation of the lenses in the cluster, one can estimate the
surface gravity of the lens giant, $g=GM/R_L^2\sim c^2/(4D)$ \citep{bro96},
given that $D$ will be of order the scale-length of the globular
cluster.

\section{Conclusions}

We have computed exact formulae for lensing of a uniform extended source 
by an opaque, spherical lens (with escape velocity much smaller than $c$).
The formulae only differ significantly from the usual occultation or 
microlensing formulae in the limit that $R_L \sim R_E$, which may be 
relevant for lensing by white dwarfs in binaries or lensing by giant 
stars.  Small deviations due to lensing in eclipsing brown-dwarf binaries 
may be detectable with very precise photometry, which may be another 
application of the expressions derived here.  A code written in 
IDL which carries out the calculations presented here can be downloaded 
from \url{http://www.astro.washington.edu/agol/}.

\acknowledgments

We thank Sara Seager for useful discussions and comments on a draft.
Support for E.A. was provided by the National Aeronautics and Space Administration 
through Chandra Postdoctoral Fellowship Award PF0-10013 issued by the Chandra 
X-ray Observatory Center, which is operated by the Smithsonian Astrophysical 
Observatory for and on behalf of the National Aeronautics Space Administration 
under contract NAS 8-39073.

\end{document}